# Chaotic diffusion in multi-scale turbulence

Yueheng Huang[1], Nong Xiang[1,*], Jiale Chen[1] and Zong Xu[2]

[1] *Institute of Plasma Physics, Chinese Academy of Sciences, Hefei 230031, China*
[2] *School of Electrical and Photoelectronic Engineering, West Anhui University, Lu'an 237012, China*

[*] *E-mail: xiangn@ipp.ac.cn*

**ABSTRACT**

This study investigates chaotic diffusion in multi-scale turbulence driven by nonlinear wave-particle resonance coupling. Turbulent waves with distinct characteristic wavelengths across scales coherently interact with charged particles when their phase velocities match the particles' velocities. A multi-wavenumber mapping framework is developed to model chaotic transport under multi-scale turbulence. By analytically deriving velocity correlation functions, we quantify the diffusion coefficient under conditions of cross-scale wave intensity parity. A critical analysis reveals that chaotic dynamics at smaller scales prove insufficient to completely erase phase-space correlations established by large-scale turbulent components. The largest-scale turbulence components dominate deviations from quasi-linear (QL) theory predictions, establishing a scale-dependent hierarchy in chaotic transport. Mere reduction of inter-wave phase velocity spacing for small-scale components cannot recover QL diffusion at finite wave amplitudes in multi-scale turbulence. Incorporating a larger-scale component into a small-scale-driven strong chaotic system can induce non-QL diffusion. Specifically, for two-scale turbulence, the QL approximation systematically underestimates transport. Increasing the number of smaller-scale components with strong overlap parameters drives convergence toward the QL approximation. This framework provides a methodology for analyzing resonance-driven turbulence in laboratory and astrophysical plasmas.

## 1. Introduction

Deciphering the nonlinear coupling between turbulence and chaos fundamentally reshapes predictive frameworks in fluid dynamics, statistical mechanics, and the study of nonlinear systems. In fusion plasmas, turbulence plays a pivotal role in energy and particle transport, exhibiting inherent multi-scale behaviour [1, 2]. Quasi-linear (QL) theory has been extensively applied to describe charged particle interactions with turbulent waves, from Langmuir turbulence saturation [3, 4] to radio-frequency wave modeling for plasma heating [5, 6] and micro-instability-driven transport [7]. However, QL diffusion validity remains temporally constrained to intervals shorter than the discretization time $\tau_d=2\pi/k_{typ}\delta v_\varphi$ ($k_{typ}$ is the typical wave number and $\delta v_\varphi$ is the interwave spacing of the wave phase velocity)[8] or requires strong resonance overlap conditions [9, 10]. Recent test particle simulations reveal nonlinear coupling between transport scales [11] , including interactions between high-frequency turbulence and low-frequency fluctuations [12] , and magnetic island-turbulence interplay [13].

While direct numerical simulation of particle orbits in multi-scale turbulence remains computationally prohibitive, mapping techniques offer powerful alternatives. The standard mapping [14, 15], particularly, provides a universal framework for studying area-preserving systems with divided phase space structure. Various dynamical systems and mappings can be locally reduced to the standard mapping, so it plays an important role in the study of classical and quantum chaos [16-18]. It has been used to model the turbulent transport of the charged particles in the wave-particle interaction[19] and more recently it was used to model the tokamak edge electron diffusion in the lower hybrid antenna electric field[20, 21]. Devations from the QL diffusion are found when the kick amplitude $K$ satisfies the Chirikov overlap condition[14], or more exactly, when $K$ is over the Greene's criterion $K_c$=0.971635…[22] in the standard mapping. For the standard mapping, the wave phases are all zeros and the ratio of the diffusion coefficient to the QL diffusion coefficient oscillates as $K$ increases and the diffusion coefficient will approach the QL diffusion limit as K very large [14, 19]. As the waves are random-phased, the diffusion coefficient never falls below the QL diffusion coefficient after rising above it[23] and self-consistent simulations of weak warm-beam instability show that the growth rate is enhanced when the diffusion coefficient is over the QL diffusion coefficient[8, 24]. Under the condition that the wavefield complex amplitudes exhibit Gaussian statistical independence, the QL approximation systematically overestimates transport[25]. The characteristic wavelengths of the turbulent waves in each of the considerations above are in same scale.

Inspired by a so-called incommensurate standard map which describes the dynamics of cold atoms in a kicked optical lattice with an incommensurate potential[26], this work extends the standard mapping formalism to multi-scale turbulence through a multi-wavenumber framework. Our analysis demonstrates that small-scale chaos cannot fully randomize phase correlations introduced by a larger scale. This finding persists across both zero-phase and random-phase mappings, with important implications for turbulent transport modeling. The article is arranged as follows: the model for the chaotic diffusion in multi-scale turbulence is presented in the second section. The analytical and numerical results of the chaotic diffusion coefficient are illustrated in the third section. Finally, the conclusions are in the last section.

## 2. Model

The electrostatic turbulent waves in one-dimensional configuration can be described as

$$E(x,t)=\sum_{m,l}E_{m,l}\sin\left(k_m x-\omega_l t+\varphi_{m,l}\right). \tag{1}$$

It is assumed that $a$ is the length of the 1-demensional configuration space, the characteristic wavenumber of the turbulent waves are $k_m=2\pi\eta_m/a$, the frequency is $\omega_l=l\omega$, where, $\eta_m$, $m$ and $l$ are integers, $\omega$ is the minimum characteristic frequency in the turbulent system, $\varphi_{m,l}$ is the wave phase. Note that if the condition $\eta_{m+1}>>\eta_m$ is given, the characteristic wavelengths $a/\eta_{m+1}$ and $a/\eta_m$ are in different scales.

We set $E_{m,l}=E_{m,0}$, so that the resonance-overlap condition for each phase velocity is the same for each $l$, for $\varphi_{m,l}=0$,

$$E(x,t)=\sum_{m=0}^{r}\sum_{l=-\infty}^{+\infty}E_{m,0}\sin\left(k_m x-l\omega t\right), \tag{2}$$

and the total number of scales is $r$+1. The normalized variables are used by rescaling the distance with $a/2\pi$, and the time with $2\pi/\omega$, the dynamics of charged particles (with the mass $\mu$ and the charge $q$) in the presence of such turbulent modes can be described as

$$\begin{cases}\dfrac{dx}{dt}=p\\[2mm] \dfrac{dp}{dt}=\displaystyle\sum_{m=0}^{r}K_m\sum_{l=-\infty}^{+\infty}\left[\sin\left(\eta_m x-2\pi l t\right)\right]\end{cases}, \tag{3}$$

where, the normalized wave amplitude $K_m=(2\pi)^3E_{m,0}q/\left(\mu a\omega^2\right)$.

Finally, a multi-wavenumber standard mapping is derived as

$$\begin{cases}p_{n+1}=p_n+\displaystyle\sum_{m=0}^{r}K_m\sin\left(\eta_m x_n\right)\\ x_{n+1}=x_n+p_{n+1}\end{cases}, \tag{4}$$

which describes multi-scale turbulence as $\eta_{m+1}>>\eta_m$.

Although $\eta_m$ may be a fractional number or even an irrational number mathematically in the mapping (4), all the $\eta_m$'s can be set as integers due to the fact that the fractional number or an irrational number can be transformed or roughly transformed to the ratio of two integers, so it can be easily set in the mapping equivalently by using integers.

Some important parameters are easily obtained. The QL diffusion coefficient of each scale of component $D_{m,QL}=\dfrac{K_m^2}{4}$. The total QL diffusion coefficient $D_{QL}=\sum_m\dfrac{K_m^2}{4}$. The interwave spacing of the wave phase velocity for each scale $\delta v_{\varphi,m}=\dfrac{2\pi}{\eta_m}$. The half-resonant width for each scale $\triangle v_m=2\sqrt{\dfrac{K_m}{\eta_m}}$, and the corresponding Chirikov overlap parameter $s_m=\dfrac{2\triangle v_m}{\delta v_{\varphi,m}}=\dfrac{2\sqrt{\eta_m K_m}}{\pi}$. Note that the overlap parameter $s_m$ with larger $\eta_m$ (smaller interwave wave phase velocity spacing $\delta v_{\varphi,m}$) increases more rapidly as $K_m$ increases than the one with smaller $\eta_m$. The Greene's criterion for each scale is $K_{m,c}=K_c/\eta_m\approx0.9716/\eta_m$. When the effects of other scales are considered, the criterion could be smaller as the Kolmogorov-Arnold-Moser surfaces may be easier broken when more perturbations of other scales are presented.

In the numerical calculation, the chaotic diffusion coefficient is measured by initiating an ensemble of 40,000 particles which are randomly distributed in configuration space with the same initial momentum $p_0$. The slope $<(p_N-p_0)^2>/(2N)$ of this ensemble gives the diffusion coefficient when $N$ is large. The method to analytically deriving velocity correlation functions of such an area-preserving map has been given in published works[27-29]. The diffusion coefficient will be shown in the third section.

## 3. Results

In this section, the calculated chaotic diffusion coefficient in multi-scale turbulence is shown. Numerical results will be used to validate the analytical calculations. For our interests, the wave amplitudes of each scale of modes are of the same order, otherwise, the terms of the lower orders can be neglected in the equation of motion. Without loss of generality, all the $K_m$'s are assumed to be the same, $K_m=K$ ($m$=0,1,…) and $\eta_0=1$ in the cases to be shown. The calculated chaotic diffusion coefficient is

$$D \approx \sum_{m=0}^{r} D_{m,QL} + \Delta_2 + \Delta_3 \, , \tag{5}$$

where, the correction term $\Delta_j$ is the phase-space-averaged impulse correlation function $\Delta_j \equiv <(p_1-p_0)\,(p_j-p_0)>$

$$\Delta_j = \sum_{m=0}^{r} D_{m,j} \left( j=2,3 \right), \tag{6}$$

the contributions to the second and third correction term of each scale is

$$D_{m,2} = -\frac{1}{4} K_m^2 \left[ \chi_2(\eta_m,0,\eta_m) + \chi_2(-\eta_m,0,-\eta_m) \right] \tag{7}$$

and

$$D_{m,3} = -\frac{K_m}{4} \sum_{t=0}^{r} K_t \begin{bmatrix} \chi_3(\eta_m,0,0,\eta_t) - \chi_3(-\eta_m,0,0,\eta_t) \\ -\chi_3(\eta_m,0,0,-\eta_t) + \chi_3(-\eta_m,0,0,-\eta_t) \end{bmatrix}, \tag{8}$$

respectively, the characteristic function $\chi_2$ and $\chi_3$ are

$$\chi_2(m_0,m_1,m_2) = \delta_{m_0,m_2} \sum_{n_0} \sum_{n_1} \cdots \sum_{n_r} \left[ \delta_{m_1+2m_2+\sum_{t=0}^{r} n_t \eta_t ,\, 0} \prod_{t=0}^{r} J_{n_t}(m_2 K_t) \right]$$

and

$$\chi_3(m_0,m_1,m_2,m_3) = \sum_{n_0} \sum_{n_1} \cdots \sum_{n_r} \left[ \chi_2 \left(m_0, m_1 - m_3, m_2 + 2m_3 + \sum_{t=0}^{r} n_t \eta_t \right) \prod_{t=0}^{r} J_{n_t}(m_3 K_t) \right]$$

respectively, and $J_n(x)$ is Bessel function of the first kind.

From Equ.(5-8), it is seen that the chaotic diffusions driven by each group of turbulent modes in different scales are coupled and they are decoupled as the correction terms are zero. To demonstrate this more clearly, we set $r$=1 which correspond to the turbulence with two scales of components. The contributions to the second and third correction terms of each scale are derived as,

$$D_{m,2} = -\frac{K_m^2}{4} \left[ {}^{(\eta_1)}J_{-2}\left(\eta_m K_0, \eta_m K_1\right) + {}^{(\eta_1)}J_{2}\left(-\eta_m K_0, -\eta_m K_1\right) \right] \tag{9}$$

and

$$D_{m,3} = -\frac{K_m}{4} \sum_{t=0}^{1} K_t \begin{bmatrix} {}^{(\eta_1)}J_{\eta_m - 2\eta_t}\left(\eta_t K_0, \eta_t K_1\right) \times {}^{(\eta_1)}J_{\eta_t - 2\eta_m}\left(\eta_m K_0, \eta_m K_1\right) \\ -{}^{(\eta_1)}J_{-\eta_m - 2\eta_t}\left(\eta_t K_0, \eta_t K_1\right) \times {}^{(\eta_1)}J_{\eta_t + 2\eta_m}\left(-\eta_m K_0, -\eta_m K_1\right) \\ -{}^{(\eta_1)}J_{\eta_m + 2\eta_t}\left(-\eta_t K_0, -\eta_t K_1\right) \times {}^{(\eta_1)}J_{-\eta_t + 2\eta_m}\left(\eta_m K_0, \eta_m K_1\right) \\ +{}^{(\eta_1)}J_{-\eta_m + 2\eta_t}\left(-\eta_t K_0, -\eta_t K_1\right) \times {}^{(\eta_1)}J_{-\eta_t + 2\eta_m}\left(-\eta_m K_0, -\eta_m K_1\right) \end{bmatrix}, \tag{10}$$

respectively, where, ${}^{(\eta)}J_n\left(x,y\right) = \sum_{q=-\infty}^{+\infty} J_{n-q\eta}\left(x\right) J_q\left(y\right)$.

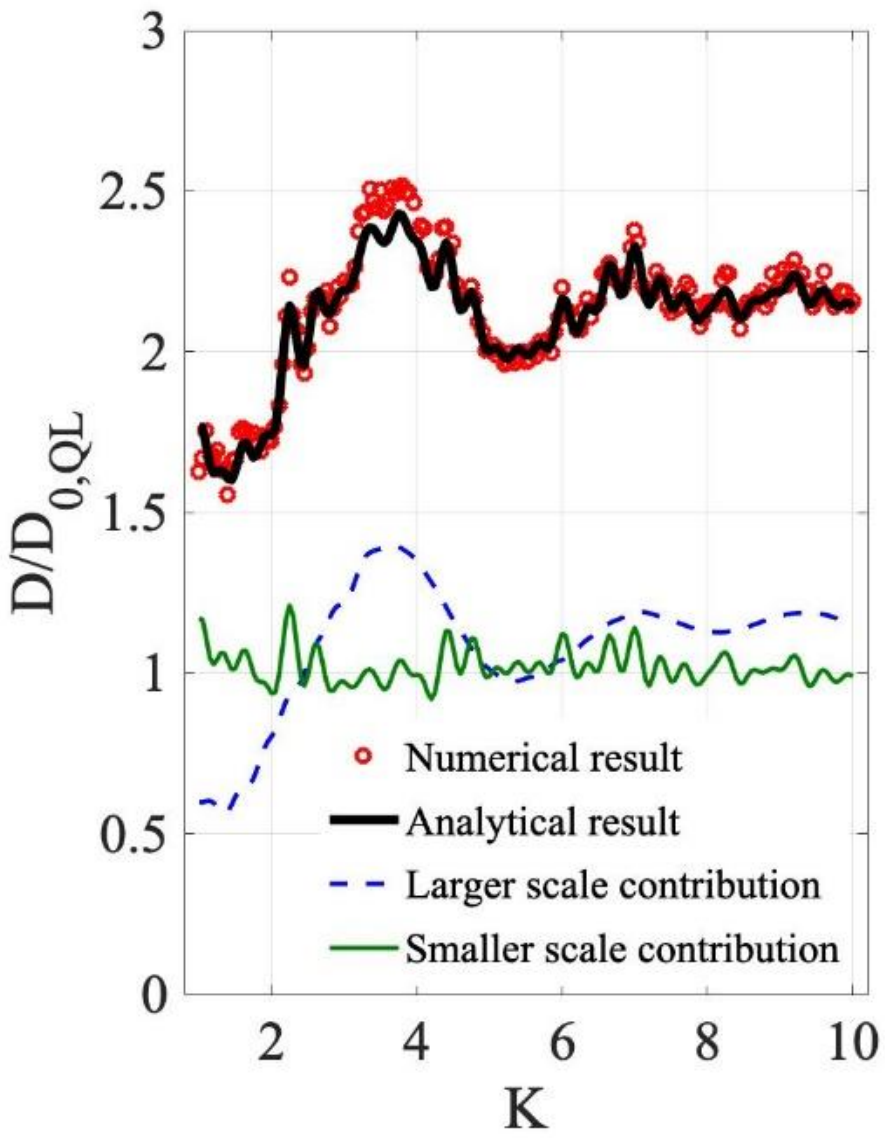

**Figure 1.** The comparison of $D/D_{0,QL}$ as a function of K between the numerical result (red circles) and theoretical result (black solid line), the larger scale contribution(blue dashed line) and smaller scale contribution(green solid line) for $r$=1, $K_0$=$K_1$=$K$, $\eta_0$ =1 and $\eta_1$ =10.

Note that the calculated diffusion coefficient in Eqs.(5-10), which is not limited to the multi-scale case, can be applied to the incommensurate standard map[26], in which the diffusion coefficient is only numerically given.

For the case with just single scale (i.e., $K_0$=0, $K_1$>1), the diffusion coefficient,

$$D \approx D_{1,QL} - \frac{K_1^2}{2} J_2(\eta_1 K_1) - \frac{K_1^2}{2}\left[ J_1^2(\eta_1 K_1) - J_3^2(\eta_1 K_1) \right] \tag{11}$$

approaches the QL diffusion coefficient as $\eta_1 K_1$ increases to a large value.

The comparison of $D/D_{0,QL}$ as a function of $K$ between the numerical result and theoretical result for $r$=1, $K_0$=$K_1$=$K$, $\eta_0$ =1 and $\eta_1$ =10, the larger scale contribution $(D_{0,QL}+D_{0,2}+D_{0,3})/D_{0,QL}$ and smaller scale contribution $(D_{1,QL}+D_{1,2}+D_{1,3})/D_{0,QL}$ are shown in Figure 1. The theoretical result agrees well with the numerical result. It is also seen that the larger scale contribution mainly contributes to the slow-varying and large amplitude oscillation as $K$ increases. It indicates that the contribution of the largest scale components dominates the correction terms of the diffusion coefficient. On the contrary, the smaller scale one mainly contribute to the fast varying and small amplitude oscillation. The fast-varying oscillation is due to the rapid increase of the overlap parameter of the smaller scale modes as $K$ increases.

However, for the multi-scale case, as the interwave spacing of the wave phase velocity of the smaller scale turbulent modes $\delta v_{\varphi,m} = \frac{2\pi}{\eta_m} \to 0$, the smaller-scale correction terms have vanished while those of the larger scale are kept (in which there is $K_m$ from smaller scale, it affect little on the oscillation amplitude as $K$ increases), the diffusion coefficient approaches a function of $K_m$'s,

$$\begin{aligned} D &\to \sum_{m=0}^{r-1} D_{m,QL} + D_{0,2} + D_{0,3} \\ &= \sum_{m=0}^{r-1} \frac{K_m^2}{4} - \frac{K_0^2}{2} J_2(K_0) \prod_{m=1}^{r} J_0(K_m) - \frac{K_0^2}{2}\left[ J_1^2(K_0) - J_3^2(K_0) \right] \prod_{m=1}^{r} J_0^2(K_m) \end{aligned} . \tag{12}$$

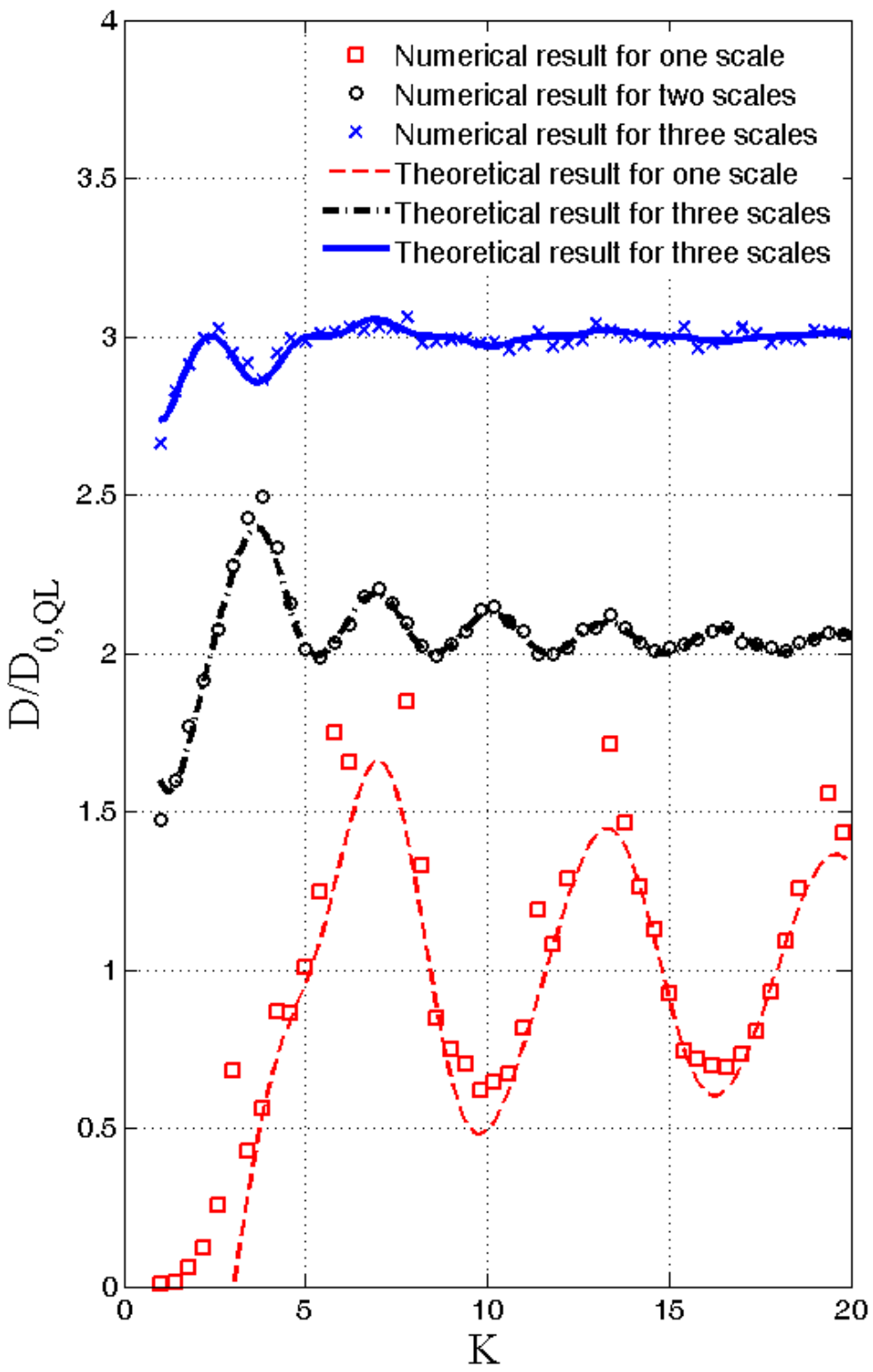

**Figure 2.** The comparison of $D/D_{0,QL}$ as a function of K between the numerical result (red squares for one scale, black circles for two scales, blue crosses for three scales) and theoretical result (red dashed line for one scale, black dash-dot line for two scales, blue solid line for three scales).

To demonstrate the characteristics of the diffusion in the large $\eta_m$ limit, three cases, i.e., turbulence including one scale of components: $r$=0, $\eta_0$=1, $K_0$=$K$, two scales of components:$r$=1, $\eta_0$=1, $\eta_1$=100, $K_0$=$K_1$=$K$, three scales of components: $r$=2, $\eta_0$=1, $\eta_1$=100, $\eta_2$=10000, $K_0$=$K_1$=$K_2$=$K$, will be compared. The theoretical expressions of $D/D_{0,QL}$ for these cases are

$$\frac{D}{D_{0,QL}} \approx 1\text{-}2J_2(K)\text{-}2\left[J_1^2(K)\text{-}J_3^2(K)\right], \tag{13}$$

which is firstly derived in Ref.[19],

$$\frac{D}{D_{0,QL}} \approx 2\text{-}2J_2(K)J_0(K)\text{-}2\left[J_1^2(K)\text{-}J_3^2(K)\right]J_0^2(K), \tag{14}$$

and

$$\frac{D}{D_{0,QL}} \approx 3\text{-}2J_2(K)J_0^2(K)\text{-}2\left[J_1^2(K)\text{-}J_3^2(K)\right]J_0^4(K), \tag{15}$$

respectively.

The comparisons of $D/D_{0,QL}$ as a function of $K$ between the numerical results and theoretical results for these three cases are shown in **Figure 2**. It is seen that the theoretical results match the numerical results well except the values at $K$~1. For $K$ ~1, more correction terms in the evaluation of Eq. (5) need to be retained. For one scale case, the theoretical result also fails to match the numerical results at $K$~$2n\pi$($n$=1,2,3,…) where the accelerator modes exist, whose existance relies on the the spacial set of the wave phases(all zero)[23]. For cases of two and three scales, the accelerator modes are not observed in our simulation because they are destroyed by the strong resonance overlap of the smaller scales. For the case of two scales, $D/D_{0,QL}$ is mostly over 2(the $QL$ diffusion limit, $D_{QL}/D_{0,QL}$=$r$, $r$=2). When $K$<2.5, $D/D_{0,QL}$ is below the QL diffusion limit in the results for the two and three scales in Figure 2. Even for $K$=1, $K$ is far beyond the Greene's criterion ($K_{m,c} \approx 0.9716/\eta_m$) for $\eta_1$=100 and $\eta_2$=10000 and just a little larger than $K_{0,c}$($\approx 0.9716/\eta_0$) for $\eta_0$=1 which indicates there are strongly overlapped resonances for the smaller scales and slightly overlapped resonances for the largest scale. The results indicate that the diffusion coefficient can deviate from the QL value when a group of turbulent waves with characteristic wavelengths at larger scale (their resonances are slightly overlapped) are added to a very chaotic system driven by turbulent waves with character wavelengths at smaller scale. From another perspective, when the QL diffusion coefficient for each scale are the same $D_{m,QL} = \frac{K^2}{4}$ while the corresponding Chirikov overlap parameter are different $s_m \propto \sqrt{\eta_m K}$ . If $K$ is finite, when the interwave spacing of the wave phase velocity $\delta v_{\varphi,m} = \dfrac{2\pi}{\eta_m} \to 0$ , $s_m \to \infty$ for the smaller scale, which indicates there is very chaotic behaved particle motion. For the larger scale, for example, $\eta_0 = 1$, $s_0 \propto \sqrt{K}$ is finite, which indicates there is not so chaotic behaved particle motion in larger scale. The results for the two-scale and three-scale cases reveal that the coupled motions results of the interplay between QL characteristic in short spatial or time scale (the smaller-scale correction terms have vanished) and non-QL characteristic in large spatial or time scale (the larger-scale correction terms are kept). So in the present of larger-scale component, the strong resonant overlap of smaller-scale turbulence proves insufficient to drive the diffusion coefficient towards the QL diffusion limit with reduced inter-wave phase velocity spacing of the smaller-scale components under the condition of finite wave amplitude. This demonstrates that small-scale turbulent component induced chaos cannot fully randomize phase correlations introduced by larger scales.

In **Figure 2,** the maximums of oscillation amplitude of $D/D_{0,QL}$ around the corresponding QL diffusion limit ($D_{QL}/D_{0,QL}=r$, $r=1,2,3$) for these three cases are >0.5,~0.5 and ~0.1, respectively, which implies that the maximum diffusion coefficient approaches the QL one as more scales of turbulent components are included. It is reasonable that as the total number of the different scales of components increases, the diffusion coefficient approaches the QL diffusion limit for $K>1$, because the correction terms are multiplied by more and more $J_0(K)$ and its absolute value is below unit as $K>1$ as shown in Eq.(13-15).

As the diffusion properties of the standard mapping is non-universal in the framework of the wave-particle interaction due to the correlated initial phases[30], a number of uncorrelated phases are used in the mapping. The randomly phased multi-wavenumber standard mapping can be written as,

$$\begin{cases} p_{n+1} = p_n + \sum_{m=0}^{r} K_m a_n \sin\left(i\eta_m x_n - \phi_{m,n}\right) \\ x_{n+1} = x_n + \dfrac{p_{n+1}}{L} \end{cases}, \tag{16}$$

where,

$$a_n e^{i\phi_{m,n}} = \frac{1}{L}\sum_{l=0}^{L-1} e^{i\left(\varphi_{m,l}+2\pi nl/L\right)}, \tag{17}$$

$\varphi_{m,l}=\varphi_{m,l+L}=random$, $L$ is the number of the different wave phases in each scale. When the total number of uncorrelated phases $L$ of each scale is large enough, the results will converge to the condition with all uncorrelated phases. Cases with the same parameters with the zero-phased mapping are used for comparison, i.e., turbulence including one scale of components: $r=0$, $\eta_0=1$, $K_0=K$, two scales of modes: $r=1$, $\eta_0=1$, $\eta_1=100$, $K_0=K_1=K$, three scales of components: $r=2$, $\eta_0=1$, $\eta_1=100$, $\eta_2=10000$, $K_0=K_1=K_2=K$. The results are shown in **Figure 3**. It is seen that the diffusion coefficient never falls below the QL diffusion coefficient after rising above it which is consistent with previous findings for one-scale case[23, 31]. For two-scale case, the diffusion coefficient is also mostly over the QL diffusion coefficient but the maximum deviation from the diffusion coefficient is lower than the one in one-scale case, which is similar with the zero-phased case. For these three cases, the maximums of deviations from the corresponding QL diffusion limit ($D_{QL}/D_{0,QL}=r$, $r=1,2,3$) are about 1.2, 0.2 and 0.1, respectively. Similar with the zero-phased case, it also confirms that the maximum diffusion coefficient approaches the QL one as more scales of turbulent components with large overlap parameters are included.

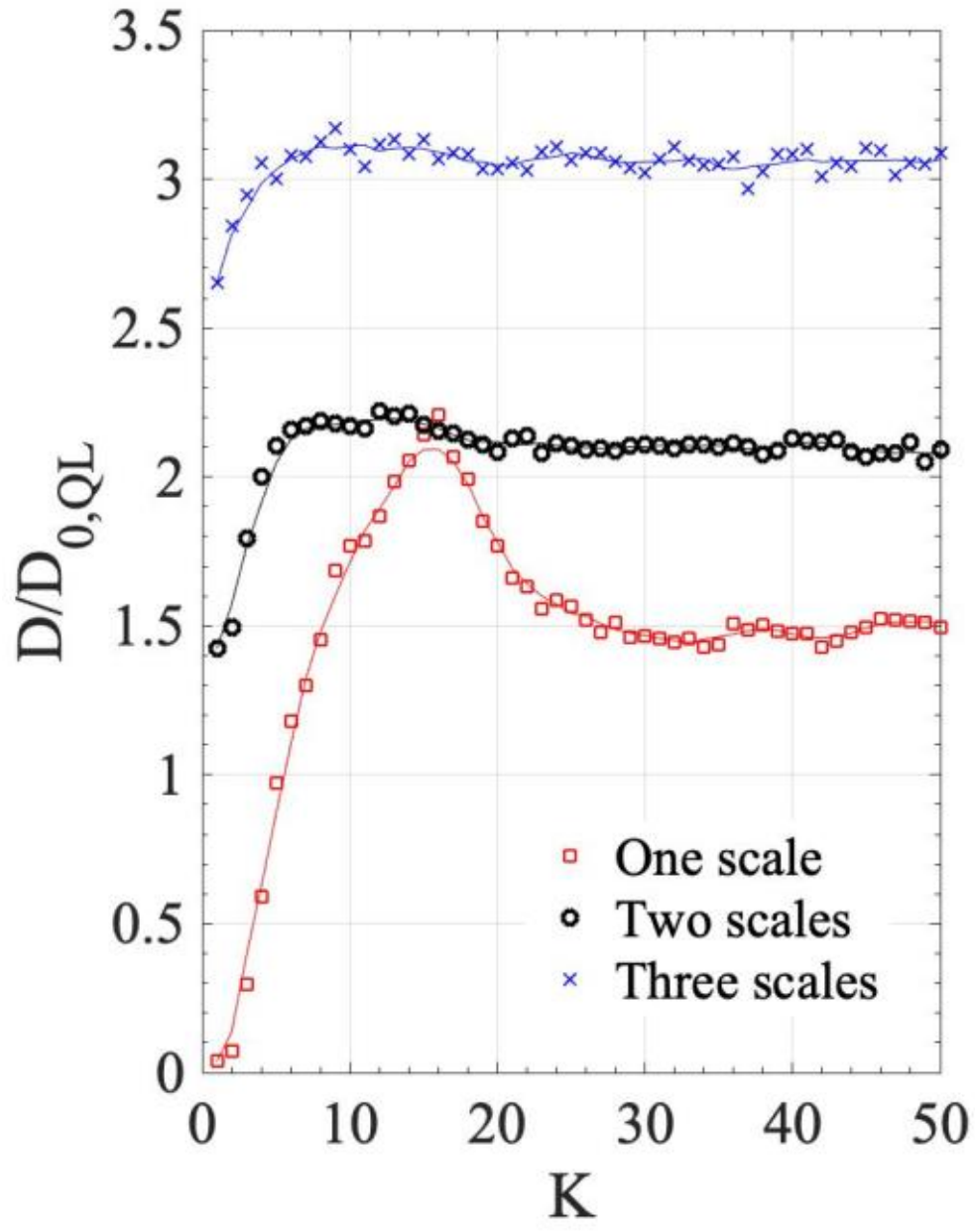

**Figure 3** The comparison of $D/D_{0,QL}$ as a function of $K$ between the numerical results of one scale (red squares), two scales (black circles), three scales (blue crosses). The lines are guides to the eyes. In the numerical calculations, $L$=50 and $N$=50$L$.

## 4. Conclusions

The multi-wavenumber standard mapping effectively models charged particle dynamics in multi-scale turbulence driven by waves with distinct characteristic wavelengths, enabling analytical derivation of chaotic diffusion coefficients under equivalent cross-scale wave intensity conditions. When resonance-overlap conditions are satisfied across scales, coupled chaotic diffusion emerges from interacting multi-scale components. Analytical calculation of velocity correlation functions reveals that deviations from quasi-linear (QL) theory predominantly originate from the largest-scale turbulence components. Both zero-phase and random-phase mappings yield consistent conclusions below.

Small-scale turbulent component induced chaos cannot fully randomize phase correlations introduced by a larger scale. Mere reduction of inter-wave phase velocity spacing for small-scale components cannot recover QL diffusion at finite wave amplitudes. Introducing larger-scale components to small-scale-driven strong chaotic systems can induce non-QL diffusion. Especially, for two-scale turbulence, the diffusion coefficient systematically exceeds QL predictions. Inclusion of additional small-scale components with strong resonance overlap drives convergence toward QL diffusion.

This framework establishes a generalized methodology for analysing resonance-driven turbulence in laboratory and astrophysical plasmas. The standard mapping's universality suggests broader applications in classical/quantum chaos studies.

## Acknowledgements

The work was supported by the Strategic Priority Research Program of the Chinese Academy of Sciences (Grant No. XDB0500302), National Nature Science Foundation under Grants Nos. 12175274, 11805133, 11905146, 12375231, 12375233 and 12275309, the Excellent Youth Project of the Education Department of Anhui Province of China (Nos. 2023AH030112). Computer resources were supported by High Performance Computing, Shenma Cluster, at Institute of Plasma Physics, Chinese Academy of Sciences.